\documentclass[submission,copyright,creativecommons]{eptcs}
\providecommand{\event}{EXPRESS/SOS 2013} 
\usepackage{breakurl}             
  
\usepackage{svn-multi}
\usepackage{amsmath}
\usepackage{amssymb}
\usepackage{verbatim} 
\usepackage{stmaryrd}
\usepackage{paralist}
\usepackage{cite}
\usepackage{appendix}
\usepackage{url}
\usepackage[]{units}
\usepackage{tikz}
\usepackage{subfig}
\usetikzlibrary{calc,arrows,positioning}
\allowdisplaybreaks

\svnid{$Id: sos.tex 1308 2013-04-25 09:50:22Z dgr240 $}

\DeclareFontFamily{U}{mathb}{\hyphenchar\font45}
\DeclareFontShape{U}{mathb}{m}{n}{
      <5> <6> <7> <8> <9> <10> gen * mathb
      <10.95> mathb10 <12> <14.4> <17.28> <20.74> <24.88> mathb12
}{}
\DeclareSymbolFont{mathb}{U}{mathb}{m}{n}
\DeclareMathSymbol{\sqdoublecup} {2}{mathb}{"5F} 
\DeclareMathSymbol{\boxplus} {2}{mathb}{"60} 

\usepackage[amsmath]{ntheorem}
\makeatletter
\renewtheoremstyle{plain}%
  {\item[\hskip\labelsep \theorem@headerfont ##1\ ##2\theorem@separator]}%
  {\item[\hskip\labelsep \theorem@headerfont ##1\ ##2, ##3\theorem@separator]}
\makeatother

\newcommand{\SOSrule}[2]{\frac{\displaystyle #1}{\displaystyle #2}}

\newcommand{\bisim}{\sim}
\newcommand{\relR}{\mathrel{\textsf{R}}}

\newcommand{\closed}[1]{{#1}\text{-closed}}

\newcommand{\DT}{\textsf{DT}}

\newcommand{\N}{\mathbb{N}} 
\newcommand{\Ninfty}{\mathbb{N}^\infty} 
\newcommand{\Q}{\mathbb{Q}} 
\newcommand{\R}{\mathbb{R}} 

\newcommand{\trans}[1][]{\xrightarrow{\, {#1} \, }}
\newcommand{\ntrans}[1][]{\mathrel{{\trans[#1]}\makebox[0em][r]{$\not$\hspace{2ex}}}{\!}}

\newcommand{\rank}{\mathop{\sf r}}
\newcommand{\openT}{\mathbb{T}}

\newcommand{\openTerms}{\openT(\Sigma)}
\newcommand{\closedTerms}{T(\Sigma)}

\newcommand{\openDT}{\mathbb{DT}}
\newcommand{\openDTerms}{\openDT(\Sigma)}

\newcommand{\Var}{\ensuremath{\textstyle{\Countvar}}}
\newcommand{\pprem}[1]{\textrm{pprem}(#1)}
\newcommand{\nprem}[1]{\textrm{nprem}(#1)}

\newcommand{\prem}[1]{\textrm{prem}(#1)}
\newcommand{\conc}[1]{\textrm{conc}(#1)}

\newcommand{\Red}[2]{\textrm{Red}}

\newcommand{\TVar}{\mathcal{V}}
\newcommand{\PVar}{\mathcal{D}}

\newcommand{\degPTSS}{\textsf{D}}

\newcommand{\Act}{A}
\newcommand{\tick}{{\surd}}

\newcommand{\ntmuft}{\ensuremath{\mathit{nt}\mkern-1.25mu \mu\mkern-1.0mu \mathit{f}\mkern-1.5mu \theta}}
\newcommand{\entmuft}{\ensuremath{\epsilon}-\ensuremath{\ntmuft}}
\newcommand{\ntmuxt}{\ensuremath{\mathit{nt}\mkern-1.5mu \mu\mkern-0.75mu \mathit{x}\mkern-0.5mu \theta}}

\newcommand{\ntmufxt}{\ensuremath{\mathit{nt}\mkern-1.25mu \mu\mkern-1.0mu \mathit{f}\mkern-1.5mu \theta\mkern-1.75mu /\mkern-1.75mu \mathit{nt}\mkern-1.5mu \mu\mkern-0.75mu \mathit{x}\mkern-0.5mu \theta}}
\newcommand{\ntyfxt}{\ensuremath{\mathit{ntyft}\mkern-1.75mu /\mkern-1.75mu \mathit{ntyxt}}}

\newcommand{\MVarMax}{\ensuremath{\textstyle{\MVar}}}

\DeclareMathOperator{\free}{free}
\DeclareMathOperator{\bound}{bound}
\DeclareMathOperator{\source}{src}
\DeclareMathOperator{\target}{trgt}
\DeclareMathOperator{\MVar}{MVar}
\DeclareMathOperator{\expbound}{exp}

\DeclareMathOperator{\Countvar}{Var}

\newcommand{\FF}{\ensuremath{F}}
\newcommand{\VV}{\ensuremath{T}}
\newcommand{\GF}{m_{\FF}}
\newcommand{\GV}{m_{\VV}}

\newcommand{\functor}{M}

\newcommand{\reactdist}[2]{\ensuremath{\textstyle{\curlyveeuparrow\!(#1,#2)}}}

\newcommand{\lfpF}{\omega_F}
\newcommand{\lfpT}{\omega_T}

\usepackage[ddmmyyyy]{datetime}
\usepackage{datenumber}
\usepackage{expl3}
\usepackage{color}

\usepackage{etex}
\reserveinserts{30}

\newenvironment{apx-proof}[1] 
        {\noindent \rm \textbf{Proof of #1.}} 
        {\qed}

\makeatletter
\newcommand*\getnumtz[2]{%
    \expandafter\@getnumtz\the\numexpr 0#2\relax
        \empty\relax\relax\@nnil{#1}{#2}%
}

\def\@getnumtz#1\relax#2\relax#3\@nnil#4#5{%
    \ifx\relax#2\relax
        \edef#4{#1}%
    \else
        \begingroup\expandafter\endgroup
        \expandafter\let\expandafter#4\csname getnumtz@#5\endcsname%
    \fi
}

\newcommand*\definetz[2]{%
    \@namedef{getnumtz@#1}{#2}%
}%

\definetz{Z}{+0000}
\definetz{GMT}{+0000}
\definetz{UTC}{+0000}
\definetz{CET}{+0100}
\definetz{CEST}{+0200}

\newcommand*\converttimezone[9]{%
    \begingroup
    \c@myyear=\numexpr#2\relax
    \c@mymonth=\numexpr#3\relax
    \c@myday=\numexpr#4\relax
    \c@myhour=\numexpr#5\relax
    \c@myminute=\numexpr#6\relax
    \c@mysecond=\numexpr#7\relax
    \getnumtz\origtz{#8}%
    \getnumtz\targettz{#9}%
    \c@myhourminute=\numexpr (#5)*100+(#6) - \origtz + \targettz \relax
    \c@myhour=\numexpr \c@myhourminute / 100\relax
    \c@myminute=\numexpr \c@myhourminute - \c@myhour*100\relax
    \loop\ifnum\c@myminute<\z@
        \advance\c@myhour by \m@ne
        \advance\c@myminute by 60\relax
    \repeat
    \loop\ifnum\c@myminute>59\relax
        \advance\c@myhour by \@ne
        \advance\c@myminute by -60\relax
    \repeat
    \ifnum\c@myhour<0\relax
        \setmydatenumber{mydatenumber}{\value{myyear}}{\value{mymonth}}{\value{myday}}%
        \advance\c@mydatenumber by \m@ne
        \setmydatebynumber{\value{mydatenumber}}{myyear}{mymonth}{myday}%
        \advance\c@myhour by 24\relax
    \else\ifnum\c@myhour>23\relax
        \setmydatenumber{mydatenumber}{\value{myyear}}{\value{mymonth}}{\value{myday}}%
        \advance\c@mydatenumber by \@ne
        \setmydatebynumber{\value{mydatenumber}}{myyear}{mymonth}{myday}%
        \advance\c@myhour by -24\relax
    \fi\fi
    \edef\@tempa{\unexpanded{#1}{\themyyear}{\themymonth}{\themyday}{\themyhour}{\themyminute}{\themysecond}{#9}}%
    \expandafter
    \endgroup\@tempa
}

\newcounter{myyear}
\newcounter{mymonth}
\newcounter{myday}
\newcounter{myhour}
\newcounter{myminute}
\newcounter{mysecond}
\newcounter{mydatenumber}
\makeatother

\ExplSyntaxOn
\cs_set_eq:NN \lgt_file_date: \today

\cs_new_nopar:Npn \lgt_file_time:
  {
    \lgt_get_time:N \currenthour :
    \lgt_get_time:N \currentminute :
    \lgt_get_time:N \currentsecond
  }

\cs_new_nopar:Npn \lgt_get_time:N #1
  {
    \int_compare:nNnT { #1 } < { 10 } { 0 }
    \int_use:N #1
  }

\ExplSyntaxOff

\definecolor{lightblue}{RGB}{224,224,255}
\definecolor{lightred}{RGB}{255,224,224}
\definecolor{lightgreen}{RGB}{224,255,224}
\definecolor{lightyellow}{RGB}{255,255,224}
\definecolor{lightpurple}{RGB}{255,224,255}
\definecolor{darkerred}{RGB}{64,0,0}
\definecolor{darkred}{RGB}{128,0,0}
\definecolor{darkblue}{RGB}{0,0,128}
\definecolor{darkgreen}{RGB}{0,128,0}
\definecolor{darkpurple}{RGB}{128,0,128}
\definecolor{grey}{rgb}{0.745098,0.745098,0.745098}
\definecolor{lightgrey}{rgb}{0.9,0.9,0.9}
\definecolor{darkgrey}{rgb}{0.6,0.6,0.6}

\makeatletter
\def\THICKhrulefill{\leavevmode \leaders \hrule height 5pt\hfill \kern \z@}
\makeatother

\newcommand{\remarkDG}[1]{}
\newcommand{\remarkST}[1]{}
\newcommand{\remarkDW}[1]{}
\newcommand{\remarkWF}[1]{}

\newcommand*\widefbox[1]{\fbox{\hspace{1em}#1\hspace{1em}}}

\usepackage{empheq}

\newtheorem{definition}{Definition}
\newtheorem{theorem}{Theorem}

\newtheorem{proposition}{Proposition}

\newenvironment{proof}{{\bf Proof. }}{$\ \ \ \ \ \ \square$\\ \smallskip}
\newcommand{\qed}{\hfill\ensuremath{\square}}
\newenvironment{example}{\ \!\!\!\!\!\!\!\!\!\!\!{\bf Example 1}}{$\hfill\ensuremath{\blacksquare}$}

\title{Compositionality of Approximate Bisimulation for Probabilistic Systems}
\author{Daniel Gebler
\institute{
	Department of Computer Science, VU University Amsterdam,\\
	De Boelelaan 1081a, NL-1081~HV~Amsterdam, The Netherlands}
\email{e.d.gebler@vu.nl}
\and 
Simone Tini
\institute{
	Department of Scienza e Alta Tecnologia, \\
	University of Insubria, Via Valleggio 11, I-22100, Como, Italy}
\email{simone.tini@uninsubria.it}
}
\def\titlerunning{Compositionality of Approximate Bisimulation for Probabilistic Systems}
\def\authorrunning{D. Gebler \& S. Tini}

\begin{document}

\maketitle

\begin{abstract}
Probabilistic transition system specifications using the rule format \ntmufxt\ provide structural operational semantics for Segala-type systems
and guarantee that probabilistic bisimilarity is a congruence. Probabilistic bisimilarity is for many applications too sensitive to the exact probabilities of transitions. Approximate bisimulation provides a robust semantics that is stable with respect to 
implementation and measurement errors of probabilistic behavior. We provide a general method to quantify how much a process combinator 
expands the approximate bisimulation distance. As a direct application we derive an appropriate rule format that 
guarantees compositionality with respect to approximate bisimilarity. Moreover, we describe how specification formats for non-standard compositionality requirements may be derived. 
\end{abstract}


\section{Introduction}\label{sec:introduction}

\remarkST{Decide the term operator/opearions/functions}
\remarkDG{Standard term in SOS literature is `function symbol' for the element of the signature. When one refers to the semantics terms in use are operators and process combinator.}
Over the last decade a number of researchers have started to develop a theory of structural operational semantics (SOS) \cite{Plo04} for probabilistic transition systems (PTSs). Several rule formats for various PTSs were proposed that ensure compositionality (in technical terms congruence) of probabilistic bisimilarity \cite{Bar04,LT05,LT09,DL12,LGD12,BM12}. The rule format \ntmufxt\cite{DL12,LGD12} subsumes all earlier formats and can be understood as the probabilistic variant of the \ntyfxt\ format \cite{Gro93}. 
Probabilistic bisimilarity is very sensitive to the exact probabilities of transitions. The slightest perturbation of the probabilities can destroy bisimilarity. Two proposals for a more robust semantics of probabilistic processes have been put forward. The \emph{metric bisimulation} approach 
\cite{GJS90,DGJP04,BW05} 
is the quantitative analogue of the relational notion of probabilistic bisimulation. It assigns a distance to each pair of processes, which measures the proximity of their quantitative properties. 
Another approach is the \emph{approximate bisimulation} (also called $\epsilon$-bisimulation) 
approach \cite{GJS90,DLT08,TDZ11}. Approximate bisimulations are probabilistic bisimulations where the transfer condition is relaxed, namely two processes are related by an $\epsilon$-bisimulation if their probability to reach a set of states related by that $\epsilon$-bisimulation differs by at most $\epsilon$.
Processes that are related by an $\epsilon$-bisimulation with $\epsilon$ being small are ``almost bisimiliar''. 
Approximate bisimulations are not transitive in general, as two states with quite different behaviors could be linked by a sequence of states, which pairwise have only little behavioral difference. Approximate bisimulations have been characterized in operational terms \cite{GJS90}, by a modal logic 
\cite{DLT08,TDZ11}, and in terms of games \cite{DLT08}. 
The metric and approximate bisimulation approach are in general not comparable (see \cite{Trac10,Tin10,TDZ11}). The main difference is that in the approximate bisimulation approach (contrary to the metric bisimulation approach) the differences along paths are neither accumulated nor weighted by the probability of the realization of that path. In this paper we consider approximate bisimulations.

In order to allow for compositional specification and reasoning, it is necessary that the considered behavioral semantics is compatible with all operators of the language of interest. For behavioral equivalences (e.g. probabilistic bisimulations) this is the well-known congruence property. For approximate bisimulations the quantitative analogue to the congruence property requires that when different processes are combined by a process combinator (i.e., an operator of the language), then the distance between the resulting combined processes 
is (reasonably) bounded. 
A natural notion for this bound is the sum of the distances between the processes to be combined \cite{DGJP04}. A process combinator respecting this specific bound is called \emph{non-expansive}. Intuitively, this bound expresses that a process combinator does not increase the behavioral distance of the processes to be combined.  The congruence property and non-expansivity property of an $n$-ary process combinator $f$ can be expressed by the following proof rules (with $\bisim$ denoting the probabilistic bisimilarity and $d$ denoting the approximate bisimulation distance):
\begin{gather*}
	\SOSrule{s_i \bisim t_i \ \text{ for all }\  i=1,\ldots,n}
			{f(s_1,\ldots,s_n) \bisim f(t_1,\ldots,t_n)}
\qquad\qquad
	\SOSrule{d(s_i,t_i) \le \epsilon_i \ \text{ for all }\  i=1,\ldots,n}
			{d(f(s_1,\ldots,s_n),f(t_1,\ldots,t_n)) \le \sum_{i=1}^n \epsilon_i}
\end{gather*}
However, for specific applications, alternative compositionality requirements are required that allow for more or less variance (than the linear sum used in non-expansivity) of the combined processes. For instance, a process combinator that combines a number of distributed systems with a measurement unit may allow for some variance in the combined distributed systems, but must enforce that the measurement unit itself is strict.

In this paper we report a substantial first step towards a theory of robust specifications for probabilistic processes. As an operational model for probabilistic processes, we consider Segala-type PTSs that exhibit both probabilistic and nondeterministic behavior. The probabilistic processes are specified by probabilistic transition system specifications (PTSS) with simple \ntmufxt\ rules. By simple \ntmufxt\ rules we mean \ntmufxt\ rules without lookahead. 
In order to facilitate compositional specification and reasoning, we study how the distance between two terms with the same topmost function symbol depends on the distances of the arguments. In detail, we characterize the \emph{expansivity} of a process combinator,
which gives an upper bound on the distance of the combined processes given the distance between their components.
Formally, the expansivity of a process combinator $f$ with $n$ arguments is defined as a mapping $\R^{n} \to \R$ taking distances of the arguments $\epsilon_1, \ldots \epsilon_n$ to $\epsilon$, with $\epsilon$ defined as the maximal distance between all $f(s_1,..,s_n)$ and $f(t_1,..,t_n)$ whenever all $s_i$ and $t_i$ are in approximate bisimulation distance $\epsilon_i$.

The first contribution of our paper is the characterization of the expansivity of each process combinator. The expansivity of a process combinator is defined as the least fixed point of a monotone function that counts recursively how often the processes are copied. Our second contribution is to deduce, from the expansivity of process combinators, an appropriate rule format that guarantees non-expansivity of all operators specified in this format. The rule format is derived from the simple \ntmufxt\ rule format by prohibiting that source processes or derivatives are copied. Finally, we demonstrate how the expansivity of process combinators can be used to derive rule formats for alternative compositionality requirements. 

We consider in this paper approximate bisimulations because the relaxed transfer condition preserves the basic relational nature of probabilistic bisimulations and allows us to apply (adapted and extended) known proof techniques developed for congruence rule formats. Moreover, the new techniques introduced in this paper to quantify the expansivity of process combinators translate naturally to bisimulation metrics. In this sense, we are also opening the door to develop a theory of robust process specifications with respect to bisimulation metrics.

This is the first paper that explores systematically the approximate bisimulation distance of probabilistic processes specified by transition system specifications.
Tini already proposed a rule format for reactive probabilistic processes \cite{Tin08,Tin10}. Our format significantly generalizes and extends that format. First of all, we apply the more general Segala-type systems that admit, besides probabilistic behavior (probabilistic choice), nondeterministic reactive behavior (internal nondeterministic branching). Furthermore, while Tini used a notion of approximate bisimulation, which is an equivalence but not closed under union, we are using the (by now) standard notion 
\cite{DLT08,TDZ11}, which is only reflexive and symmetric but closed under union. Finally, the novel rule format 
based on counting of copies of processes and their derivatives in its defining rules allows us to handle a wider class of process combinators that ensure non-expansivity.
\section{Preliminaries}\label{sec:preliminaries}

We assume an infinite set of (state) variables $\TVar$. We let $x, y, z$ range over $\TVar$. A \emph{signature} is a structure $\Sigma = (F, \rank)$, where 
\begin{inparaenum}[(i)]
	\item $F$ is a set of \emph{function names} (operators) disjoint from $\TVar$, and
	\item $\rank : F \to \N$ is a \emph{rank function}, which gives the arity of a function name. An operator $f \in F$ is called a \emph{constant} if $\rank(f)=0$.
\end{inparaenum}
We write $f\in\Sigma$ for $f\in F$.  
Let $W \subseteq \TVar$ be a set of variables. The set of $\Sigma$-terms (also called state terms) over $W$, denoted by $T(\Sigma, W)$, is the least set satisfying: 
\begin{inparaenum}[(i)]
	\item $W \subseteq T(\Sigma, W)$, and
	\item if $f \in \Sigma$ and $t_1, \cdots, t_{\rank(f)} \in T(\Sigma, W)$, then $f (t_1, \cdots, t_{\rank(f)}) \in T(\Sigma, W)$.
\end{inparaenum}
$T(\Sigma, \emptyset)$ is the set of all \emph{closed terms} and abbreviated as $\closedTerms$. $T(\Sigma, \TVar)$ is the set of \emph{open terms} and abbreviated as $\openTerms$. We may refer to operators as process combinators, and refer to terms as processes. $\Var(t) \subseteq \TVar$ denotes the set of variables in $t$. 
$\MVar\!: \openTerms \to (\TVar \to \N)$ denotes for $\MVar(t)(x)$ how often the variable $x$ occurs in $t$.
A (state variable) \emph{substitution} is a mapping $\sigma_\TVar: \TVar \to \openTerms$.
A substitution is closed if it maps each variable to a closed term. A substitution extends to a mapping from terms to terms as usual.
\remarkWF{Should the substitution not be better a partial map?}
\remarkDG{Total function is easier to handle when later defining closed instantiations of rules (otherwise one always needs to say that variables are replaced by closed terms AND the domain of the substitution is at least the set of variables of a rule)}

Let $\Delta(\closedTerms)$ denote the set of all (discrete) probability distributions on $\closedTerms$. We let $\pi$ range over $\Delta(\closedTerms)$. For $S \subseteq \closedTerms$ we define $\pi(S)=\sum_{t\in S}\pi(t)$. For each $t \in \closedTerms$, let $\delta_t$ denote the \emph{Dirac distribution}, i.e., $\delta_t(t)=1$ and  $\delta_t(t')=0$ if $t$ and $t'$ are not syntactically equal. The convex combination $\sum_{i \in I}p_i \pi_i$ of a family $\{\pi_i\}_{i \in I}$ of probability distributions with $p_i \in (0,1]$ and $\sum_{i \in I} p_i = 1$ is defined by $(\sum_{i \in I}p_i \pi_i)(t) = \sum_{i \in I} (p_i \pi_i(t))$. 
By $f(\pi_1,\dots,\pi_{\rank(f)})$ we denote the distribution that is defined by $f(\pi_1,\dots,\pi_{\rank(f)})(f(t_1,\ldots,t_{\rank(f)})) = \prod_{i=1}^{\rank(f)}\pi_i(t_i)$. We may use the infix notation where appropriate.

In order to describe probabilistic behavior, we need expressions that denote probability distributions. We assume an infinite set of distribution variables $\PVar$. We let $\mu$ range over $\PVar$ and $\zeta$ range over $\PVar \cup \TVar$. Let $D \subseteq \PVar$ be a set of distribution variables and $V\subseteq\TVar$ be a set of state variables.  The set of \emph{distribution terms} over $D$ and $V$, notation $\DT(\Sigma, D, V)$, is the least set satisfying: 
\begin{inparaenum}[(i)]
	\item \label{def:DT:var_and_inst_dirac}
		$D \cup \{\delta_t \mid t\in T(\Sigma, V)\} \subseteq \DT(\Sigma, D, V)$, 
	\item \label{def:DT:sum} 
		${\textstyle \sum_{i\in I} p_i \theta_i \in \DT(\Sigma, D, V)}$ if $\theta_i\in \DT(\Sigma, D, V)$ and $p_i \in (0,1]$ with $\sum_{i\in I} p_i = 1$, and
	\item \label{def:DT:prod} 
		$f(\theta_1,\ldots,\theta_{\rank(f)}) \in \DT(\Sigma, D, V)$ if $f\in\Sigma$ and $\theta_i\in \DT(\Sigma, D, V)$.\footnote{This fixes a flaw in \cite{LGD12,DL12} where arbitrary functions $f\!\!:\closedTerms^n \to \closedTerms$ were allowed. In this case probabilistic bisimilarity (Definition~\ref{def:bisimulation}) may not be a congruence (Theorem~\ref{thm:bisimilarity_as_congruence}). Example: PTSS $(\Sigma,A,R)$, constants $r,r',s$ in $\Sigma$, $A=\{a\}$, $R=\left\{\SOSrule{}{s \trans[a] \delta_s}, \SOSrule{x \ntrans[a]}{g(x) \trans[a] f(\delta_x)}\right\}$ with $f(r)=r, f(r')=s$. Now $r \bisim r'$ but $g(r) \not\bisim g(r')$.} 
\end{inparaenum}
A \emph{distribution variable} $\mu\in D$ is a variable that takes values from $\Delta(\closedTerms)$. An \emph{instantiable Dirac distribution} $\delta_t$ with $t\in\openTerms$ is a symbol that takes value $\delta_{t'}$ when variables in $t$ are substituted so that $t$ becomes the closed term $t'\in\closedTerms$. Case \ref{def:DT:sum} allows one to construct convex combinations of distributions. For concrete terms we use the infix notation, e.g., $[p_1]\theta_1 \oplus [p_2]\theta_2$ for $\theta = \sum_{i\in \{1,2\}} p_i\theta_i$. Case \ref{def:DT:prod} lifts the structural inductive construction of state terms to distribution terms.
$\DT(\Sigma, \PVar, \TVar)$ is abbreviated as $\openDTerms$.


$\MVar\!: \openDTerms \to (\TVar \cup \PVar \to \N)$ denotes for $\MVar(\theta)(\zeta)$ how often the variable $\zeta$ occurs in $\theta$. For convex combinations $\sum_{i\in I} p_i \theta_i$ the maximal occurrence in some $ \theta_i$ is considered because the probabilistic choice selects (probabilistically) exactly one of the summands.
Formally, we have $\MVarMax(\mu)(\mu) = 1$, $\MVarMax(\mu)(\zeta) = 0$ if $\mu\neq\zeta$, $\MVarMax(\delta_t)(x) = \MVar(t)(x)$, $\MVarMax(\delta_t)(\mu) = 0$, $\MVarMax(\sum_{i\in I} p_i \theta_i)(\zeta) = \max_{i\in I} \MVarMax(\theta_i)(\zeta)$, and $\MVarMax(f(\theta_1,\ldots,\theta_{\rank(f)}))(\zeta)$ = $\sum_{i=1}^{\rank(f)} \MVarMax(\theta_i)(\zeta)$. 
A substitution on state and distribution variables is a mapping $\sigma: (\TVar \cup \PVar) \to (\openTerms \cup \openDTerms)$ such that $\sigma(x) \in \openTerms$ if $x\in \TVar$, and $\sigma(\mu) \in \openDTerms$ if $\mu\in \PVar$. A substitution extends to distribution terms by $\sigma(\delta_t)=\delta_{\sigma(t)}$, $\sigma(\sum_{i\in I} p_i \theta_i) = \sum_{i\in I} p_i \sigma(\theta_i)$ and $\sigma(f(\theta_1,\ldots,\theta_{\rank(f)})) = f(\sigma(\theta_1),\ldots,\sigma(\theta_{\rank(f)}))$. Notice that closed instances of distribution terms are probability distributions. 

\section{Probabilistic Transition System Specifications}\label{sec:ptss}

Probabilistic transition systems (PTSs) generalize labelled transition systems (LTSs) by allowing for probabilistic choices in the transitions. 
We consider nondeterministic PTSs (Segala-type systems) \cite{Seg95a} with countable state spaces.

\begin{definition}{\ \!\!\!\emph{(PTS)}\bf{.}\,}
A \emph{probabilistic labeled transition system} (PTS) is a triple $(\closedTerms,\Act,{\trans})$, where $\Sigma$ is a signature, $\Act$ is a countable set of actions, and ${\trans} \subseteq \closedTerms \times \Act \times \Delta(\closedTerms)$ is a transition relation.
\end{definition}
We write $s \trans[a] \pi$ for ${(s,a,\pi)} \in {\trans}$. PTSs are specified by means of transition system specifications \cite{Plo04,Gro93,GV92,LGD12}.

\begin{definition}{\ \!\!\!\emph{(Simple \ntmufxt-rule)}\bf{.}\,} \label{def:simple_ntmuft}
A \emph{simple \ntmuft-rule} has the form:
\[
	\SOSrule{\{  t_k \trans[a_k] \mu_k \mid k \in K\} \qquad
			 \{ {t_l \ntrans[b_l]} \mid {l \in L}\} }
			{f(x_1,\ldots,x_{\rank(f)}) \trans[a] \theta}
\]
with $t_k,t_l\in \openTerms$, $a_k,b_l,a \in \Act$, $\mu_k\in\PVar, f\in\Sigma,x_1,\ldots,x_{\rank(f)}\in\TVar$, $\theta\in\openDTerms$, and constraints:
\begin{enumerate}
	\item \label{cond:simple_ntmufxt:pairwise_difference_positive_literal} 
		all $\mu_k$ for $k \in K$ are pairwise different; 
	\item \label{cond:simple_ntmufxt:pairwise_difference_source}
		all $x_1,\ldots,x_{r(f)}$ are pairwise different.
\end{enumerate}
A \emph{simple \ntmuxt-rule} is as above with source of its conclusion $x\in\TVar$ instead of $f(x_1,\ldots,x_{\rank(f)})$.
A \emph{simple \ntmufxt-rule} is either a simple \ntmuft-rule or a simple \ntmuxt-rule. 
\end{definition}
%
The expressions $t_k \trans[a_k] \mu_k$ (resp.\ $t_l \ntrans[b_l]$) above the line are called \emph{positive} (resp. \emph{negative}) \emph{premises}. We call $\mu_k$ in $t_k \trans[a_k] \mu_k$ a \emph{derivative} for each $x \in \Var(t_k)$. For rule $\rho$ we denote the set of positive (resp.\ negative) premises by $\pprem{\rho}$ (resp.\ $\nprem{\rho}$), and the set of all premises by $\prem{\rho}$.
A rule without premises is called an \emph{axiom}. We allow 
the sets of positive and negative premises to be infinite.
The expression $f(x_1,\ldots,x_{\rank(f)}) \trans[a] \theta$ below the line is called \emph{conclusion}, notation $\conc{\rho}$. The term $f(x_1,\ldots,x_{\rank(f)})$ is called the \emph{source} of $\rho$, notation $\source(\rho)$, and $x_i$ are the \emph{source variables}, notation $x_i \in \source(\rho)$. $\theta$ is the \emph{target} of $\rho$, notation $\target(\rho)$.
An expression $t \trans[a] \theta$ (resp.\ $t \ntrans[a]$) is called a \emph{positive} (resp.\ \emph{negative}) \emph{literal}.
Hence, premises and conclusions are literals. We denote the set of variables in $\rho$ by $\Var(\rho)$, \emph{bound variables} by $\bound(\rho) = \{x_1,\dots,x_{\rank(f)}\} \cup \{\mu_k \mid k\in K\}$, and \emph{free variables} by $\free(\rho) = \Var(\rho) \setminus \bound(\rho)$.

A \emph{probabilistic transition system specification} (PTSS) in simple \ntmufxt-format, called simple \ntmufxt-PTSS for short, is a triple $P = (\Sigma,A,R)$ with $\Sigma$ a signature, $\Act$ a set of action labels, and $R$ a set of simple \ntmufxt-rules. 

As PTSS have negative premises, there are multiple approaches to assign a meaning (see \cite{vG04} for an overview). We will use the stratification approach presented in \cite{DL12} to assign to each PTSS $P = (\Sigma,\Act,R)$ (if possible) a PTS $(\closedTerms,\Act,{\trans}_P)$. A closed literal $t \trans[a]\pi$ (resp.\ $t \ntrans[a]$) \emph{holds in} ${\trans}_P$, notation ${{\trans}_P} \models t \trans[a]\pi$ (resp.\ ${{\trans}_P} \models t \ntrans[a]$), if $(t, a, \pi) \in {{\trans}_P}$ (resp.\ there is no $\pi \in \Delta(\closedTerms)$ s.t.\ $(t, a, \pi) \in {{\trans}_P}$).
A substitution $\sigma$ extends to literals by $\sigma(t \trans[a] \mu) = \sigma(t) \trans[a] \sigma(\mu)$, and $\sigma(t \ntrans[a]) = \sigma(t) \ntrans[a]$, and to rules as expected.

\begin{definition}{\ \!\!\!\emph{(Stratification \cite{DL12})}\bf{.}\,} \label{def:stratification}
Let $P = (\Sigma, \Act, R)$ be a PTSS.  A function $S:\closedTerms \times \Act \times \Delta(\closedTerms) \to \alpha$, where $\alpha$ is an ordinal, is called a \emph{stratification} of $P$ if for every rule $\rho$
\[
	\SOSrule{\{  t_k \trans[a_k] \mu_k \mid k \in K\} \quad
			 \{ {t_l \ntrans[b_l]} \mid {l \in L}\} 
			}
			{f(x_1,\ldots,x_{\rank(f)}) \trans[a] \theta}
\]
in $R$ and substitution $\sigma : (\TVar\cup\PVar) \to (\closedTerms\cup\Delta(\closedTerms))$ 
we have:
\begin{inparaenum}[(i)]
	\item $S(\sigma(t_k \trans[a_k] \mu_k)) \leq S(\conc{\sigma(\rho)})$ for all $k\in K$, and
	\item $S(\sigma(t_l \trans[b_l] \mu)) < S(\conc{\sigma(\rho}))$ for all $l\in L, \mu \in \PVar$.
  \end{inparaenum}
The set $S_{\beta} = \{\psi \mid S(\psi)=\beta\}$, with $\beta<\alpha$, is called a \emph{stratum}. 
\end{definition}
We call $P$ \emph{stratifiable} if $P$ has some stratification. A transition relation is constructed stratum by stratum in an increasing manner.
%
\begin{definition}{\ \!\!\!\emph{(Induced PTS \cite{DL12})}\bf{.}\,} \label{def:pts_induced_by_ptss}
Let $P = (\Sigma, \Act, R)$ be a PTSS with stratification $S:\closedTerms \times \Act \times \Delta(\closedTerms) \to \alpha$. For all rules $\rho$, let $\degPTSS(\rho)$ be the smallest regular cardinal greater than $|\pprem{\rho}|$, and let $\degPTSS(P)$ be the smallest regular cardinal such that $\degPTSS(P)\geq\degPTSS(\rho)$ for all $\rho\in R$. The \emph{induced PTS} $(\closedTerms, \Act, \trans_{P, S})$ 
is defined by ${\trans_{P,S}} = \bigcup_{\beta < \alpha} {\trans_{P_\beta}}$, where $\trans_{P_\beta} = \bigcup_{j \leq \degPTSS(P)} {\trans_{P_{\beta,j}}}$ and ${\trans_{P_{\beta,j}}}$ is 
\begin{align*}
	{\trans_{P_{\beta,j}}} = 
		\left\{ \psi \,\left|\,
		\begin{array}{l}
			 S(\psi) = \beta \text{ and } \exists \rho \in R \text{ and substitution } \sigma \text{ s.t. } \psi = \conc{\sigma(\rho)}, \text{ and} \\[.7ex]
		     \textstyle(\bigcup_{ \gamma < \beta} {\trans_{P_\gamma}}) \cup (\bigcup_{ j' < j} {\trans_{P_{\beta,j'}}}) 
\models {\pprem{\sigma(\rho)}}, \text{ and} \\[.7 ex]
		     \textstyle(\bigcup_{ \gamma < \beta} {\trans_{P_\gamma}}) \models \nprem{\sigma(\rho)} 
		\end{array}
		\right.\right\}
\end{align*}
\end{definition}
%
The induced PTS is independent from the chosen stratification \cite{DL12}. We can construct for each simple \ntmufxt-PTSS $(\Sigma, \Act, R)$ a PTSS $(\Sigma, \Act, R')$ with only simple \ntmuft-rules that induces the same PTS \cite{LGD12}. The construction defines $R'$ as $R$ where each rule with a source of the form $x$ is replaced by a set of rules where $x$ is substituted by $f(x_1,\ldots,x_{\rank(f)})$ for each $f\in \Sigma$ . Hence, all our results below for simple \ntmuft-PTSS generalize to simple \ntmufxt-PTSS. 


Given a relation ${\relR} \subseteq \closedTerms \times \closedTerms$, a set $X \subseteq \closedTerms$ is $\closed{\relR}$,
denoted by $\closed{\relR}(X)$, if ${\relR}(X) \subseteq X$ where ${\relR}(X) = \{y \in \closedTerms \mid \exists x \in X. x \relR y\}$.

\begin{definition}{\ \!\!\!\emph{(Probabilistic Bisimulation \cite{LS91,DGJP03})}\bf{.}\,} \label{def:bisimulation}
Let $(\closedTerms,\Act,{\trans})$ be a PTS. A symmetric relation ${\relR} \subseteq \closedTerms \times \closedTerms$ is a \emph{probabilistic bisimulation} if whenever $t \relR t'$ and $t \trans[a] \pi$ then there exists a transition $t' \trans[a] \pi'$  such that $\pi \relR \pi'$, where
\[
	\pi \relR \pi' \quad\text{iff}\quad\text{for all } X\subseteq \closedTerms \text{ with } \closed{\relR}(X) \text{ we have } \pi(X) = \pi'(X).
\]
\end{definition}
Notice that this standard definition can be slightly reformulated to relate it to the later introduced $\epsilon$-bisimulation (Definition~\ref{def:epsilon_bisim}) by requiring that $\pi \relR \pi'$ iff $\pi(X) \le \pi'({\relR}\,(X))$ for all $X\subseteq \closedTerms$ \cite{DLT08}. The union of all probabilistic bisimulations is the largest probabilistic bisimulation, called probabilistic bisimilarity, and denoted by $\bisim$. 
We shall refer to probabilistic bisimulation as strict bisimulation to distinguish it from the later introduced relaxed notion of $\epsilon$-bisimulation.

A crucial property of process description languages to ensure compositional modelling and verification is the compatibility of process operators with the behavioral relation chosen for the application context. In algebraic terms the compatibility of a behavioral equivalence $\relR$ with operator $f\in \Sigma$ is expressed by the congruence property which is defined as $f(t_1,\ldots,t_{\rank(f)}) \relR f(u_1,\ldots,u_{\rank(f)})$ whenever $t_i \relR u_i$ for $i=1,\ldots,\rank(f)$. The rule format of Definition~\ref{def:simple_ntmuft} is an instance of the \ntmufxt\ rule format \cite{LGD12}, which ensures that bisimilarity is a congruence.

\begin{theorem}{\ \!\!\!\emph{(Probabilistic Bisimilarity as a congruence \cite{LGD12})}\bf{.}\,} \label{thm:bisimilarity_as_congruence}
Let $P=(\Sigma,\Act,R)$ be a stratifiable simple \ntmufxt-PTSS. Then probabilistic bisimilarity is a congruence for all operators defined in $P$.
\end{theorem}
In order to allow for robust reasoning on PTSs, the behavioral relations 
should allow for (limited) perturbation of probabilities \cite{GJS90}. $\epsilon$-bisimulation is a behavioral relation based on strict probabilistic bisimulation, where the transfer condition is relaxed by some upper bound on the pertubation of probabilities. 

\begin{definition}{\ \!\!\!\emph{($\epsilon$-Bisimulation \cite{DLT08})}\bf{.}\,} \label{def:epsilon_bisim} 
Let $(\closedTerms,\Act,{\trans})$ be a PTS and $\epsilon \in [0,1]$. A symmetric relation ${\relR} \subseteq \closedTerms \times \closedTerms$ is an \emph{$\epsilon$-bisimulation} if whenever $t \relR t'$ and $t \trans[a] \pi$ then there exists a transition $t' \trans[a] \pi'$  such that $\pi \relR \pi'$, where
\[
	\pi \relR \pi' \quad\text{iff}\quad \text{for all } X\subseteq \closedTerms \text{ we have } \pi(X) \le \pi'({\relR}\,(X)) + \epsilon.
\]
\end{definition}
We call $t$ and $t'$ (resp.\ $\pi$ and $\pi'$) $\epsilon$-bisimilar if $t \relR t'$ (resp.\ $\pi \relR \pi'$) for some $\epsilon$-bisimulation $\relR$. Notice that $\epsilon$-bisimulations are reflexive and symmetric but 
not necessarily transitive. 
$\epsilon$-bisimulations are closed under union. We denote the largest $\epsilon$-bisimulation, called $\epsilon$-bisimilarity, by $\bisim_\epsilon$.  
According to \cite{DLT08}, $\epsilon$-bisimulations induce a pseudo-metric over the set of closed terms $d:\closedTerms \times \closedTerms \rightarrow [0,1]$ with $d(t,t') = \inf\{\epsilon\in[0,1] \mid t \bisim_{\epsilon} t'\}$, where $\inf\, \emptyset = 1$. We say that $t$ and $t'$ are within the approximation bisimulation distance $\epsilon$ if $d(t,t')=\epsilon$.



\section{Expansivity of Process Combinators}\label{sec:aptss}

The expansivity of an operator $f \in \Sigma$ is defined as the maximal approximate bisimulation distance of terms with an outermost function symbol $f$ in relation to the approximate bisimulation distances of its arguments.
In this section we quantify the expansivity of operators defined by a PTSS. We start by showing that 
the expansivity of an operator $f$ defined by a rule $\rho$ depends on
\begin{inparaenum}[(i)]
	\item the \emph{multiplicity} (i.e. number of occurrences) of source variables and their derivatives in the target of $\rho$; 
	\item the \emph{expansivity power} of operators (i.e. how much does the operator multiply the distance of its arguments) that define a context 
around the source variables or their derivatives;
and
	\item the (reactive behavior) \emph{discriminating power} of the premises of $\rho$.
\end{inparaenum}


\begin{example}{\,(Factors of Expansivity)\bf{.}\,} \label{ex:expansivity_of_operators_approx_bisim}
Let $(\Sigma, \Act, R)$ be a PTSS with a signature $\Sigma$ that contains constants $r,s,0$, unary function symbols $f,f_2$, binary function symbols $g,g_2,g_3$ and a quaternary function symbol $h$, action set $\Act=\{a\}$, and axioms $R = \{r \trans[a] \delta_r, s \trans[a] [1-\epsilon]\delta_s \oplus [\epsilon]\delta_0\}$ for some fixed $\epsilon\in(0,1)$. It is not hard to see that $d(r,s)=\epsilon$ in the PTS induced by $(\Sigma,\Act,R)$. Consider the rules:
\begin{gather*}
	\SOSrule{x \trans[a] \mu}{f(x) \trans[a] g(\mu,\mu)}
\qquad\qquad
	\SOSrule{x_1 \trans[a] \mu_1 \quad x_2 \trans[a] \mu_2}{g(x_1,x_2) \trans[a] g(\mu_1,\mu_2)}
\end{gather*}
These rules together with $R$ define $R_2$. In the first rule the derivative $\mu$ of source variable $x$ appears twice in the rule target $g(\mu,\mu)$. The induced PTS of  $(\Sigma,\Act,R_2)$ contains the following transitions:

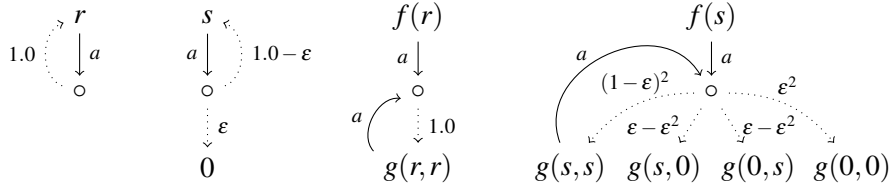
\vspace{0.2cm}\hspace{1cm}
\begin{tikzpicture}
	\node (r) at (0,0) {$r$} ;
	\node (pir) at ($ (r) - (0,1) $) {$\circ$};
	\path[->] (r)  edge node [right] {{\scriptsize $a$}} (pir);
	\path[->] (pir) [bend left = 70,dotted] edge node [left] {{\scriptsize $1.0$}} (r);

	\node (s) at ($ (r) + (1.7,0) $) {$s$} ;
	\node (pis) at ($ (s) - (0,1) $) {$\circ$};
	\node (s') at ($ (pis) - (0,1) $) {$0$};

	\path[->] (s)  edge node [left] {{\scriptsize $a$}} (pis);
	\path[->] (pis) [bend right = 70,dotted]  edge node [right] {{\scriptsize $1.0-\epsilon$}} (s);		
	\path[->] (pis) [dotted] edge node [right] {{\scriptsize $\epsilon$}} (s');

	\node (r) at ($ (s) + (2.8,0) $) {$f(r)$} ;
	\node (pir1) at ($ (r) - (0,1) $) {$\circ$};
	\node (r') at ($ (pir1) - (0,1) $) {$g(r,r)$} ;

	\path[->] (r) edge node [left] {{\scriptsize $a$}} (pir1);
	\path[->] (pir1) [dotted] edge node [right] {{\scriptsize $1.0$}} (r');
    \path[->] (r') [bend left = 70] edge node [left] {{\scriptsize $a$}} (pir1);

	\node (fs) at ($ (r) + (3.9,0) $) {$f(s)$} ;
	\node (pis) at ($ (fs) - (0,1) $) {$\circ$};
	\node (hs1) at ($ (pis) - (1.875,1) $) {$g(s,s)$} ;
	\node (hs2) at ($ (pis) - (0.625,1) $) {$g(s,0)$} ;
	\node (hs3) at ($ (pis) - (-0.625,1) $) {$g(0,s)$} ;
	\node (hs4) at ($ (pis) - (-1.875,1) $) {$g(0,0)$} ;
	
	\path[->] (fs) edge node [right] {{\scriptsize $a$}} (pis);
	\path[->] (pis) [dotted, bend right = 25] edge node [above] {{\scriptsize $(1-\epsilon)^2$}} (hs1);
	\path[->] (pis) [dotted] edge node [left] {{\scriptsize $\epsilon - \epsilon^2$}} (hs2);
	\path[->] (pis) [dotted] edge node [right] {{\scriptsize $\epsilon - \epsilon^2$}} (hs3);
	\path[->] (pis) [dotted, bend left = 25] edge node [above] {{\scriptsize $\epsilon^2$}} (hs4);
	\path[->] (hs1) [bend left = 85,distance=1.05cm] edge node [left,xshift=-0.1cm] {{\scriptsize $a$}} (pis);
\end{tikzpicture}
\vspace{0.1cm}

\noindent
Observe that $d(f(r),f(s))=1-(1-\epsilon)^2$. The power of $2$ in the distance reflects directly the multiplicity of $2$ of the derivative $\mu$ in the rule target. 
The same effect can be observed for multiple occurrences of source variables in the rule target, e.g. consider for the $f$-defining rule $g(\delta_x,\delta_x)$ instead of $g(\mu,\mu)$ as target.

Furthermore, the expansivity power of operators used in the rule target determine the expansivity of the operator defined by that rule. A simple example is the axiom $f_2(x) \trans[a] \delta_{f(x)}$. While the variable $x$ occurs only once in the rule target,
we still have $d(f_2(r),f_2(s))=1-(1-\epsilon)^2$, because the operator $f$ 
has an expansivity power of 2 
wrt.\ its single argument. This indicates that the expansivity power of (arguments of) operators need to be defined recursively.

The multiplicity of source variables and their derivatives and the expansivity power of operators applied on those variables multiply. Consider the rules:
\begin{gather*}
	\SOSrule{x_1 \trans[a] \mu_1 \quad x_2 \trans[a] \mu_2}{g(x_1,x_2) \trans[a] h(\mu_1,\mu_1,\mu_2,\mu_2)}
\qquad\qquad
	\SOSrule{\{x_i \trans[a] \mu_i \mid i=1,\ldots,4\}}{h(x_1,x_2,x_3,x_4) \trans[a] h(\mu_1,\mu_2,\mu_3,\mu_4)}
\end{gather*}
These rules together with $R_2$ define $R_3$. 
\noindent
Now $d(f(r),f(s))=1-(1-\epsilon)^4$. As explained above for $R_2$, in the rule defining operator $f$ the derivative $\mu$ appears twice in the rule target. Additionally the operator $g$ that is applied to $\mu$ has for both of its arguments an expansivity power of two because in the $g$-defining rule the derivatives $\mu_1, \mu_2$ of both arguments $x_1, x_2$ appear twice in the rule target. 

The expansivity power of an operator may be unbounded. Consider the recursive unary operator $f$ defined by the rules:
\begin{gather*}
	\SOSrule{x \trans[a] \mu}{f(x) \trans[a] g(f(\mu),f(\mu))}
\qquad\qquad
	\SOSrule{x_1 \trans[a] \mu_1 \quad x_2 \trans[a] \mu_2}{g(x_1,x_2) \trans[a] g(\mu_1,\mu_2)}
\end{gather*}
These rules together with $R$ define $R_4$. In the rule that defines the operator $f$ the derivative $\mu$ occurs twice in the target. Moreover, each occurrence of $\mu$ is put in the context of that operator $f$, which is defined by this rule (recursive call). Additionally both occurrences of $f(\mu)$ are put in the binary context $g$, which enforces that the distances of the two copies of $\mu$ multiply.
Recursive multiplication of the distances leads to an approximate bisimulation distance of $d(f(r),f(s)) = 1$. 
The expansivity power of $f$ will in this case be denoted by $\infty$. 

On the other hand, an operator may also absorb the approximate bisimulation distance.
Consider the rules:
\begin{gather*}
	\SOSrule{x \trans[a] \mu}{f(x) \trans[a] g_2(\mu,\mu)}
\qquad\qquad
	\SOSrule{x \trans[a] \mu}{f_2(x) \trans[a] g_3(\mu,\mu)}
\qquad\qquad
	\SOSrule{}{g_3(x_1,x_2) \trans[a] \delta_0}
\end{gather*}
These rules together with $R$ define $R_5$. The first rule applies the undefined operator $g_2$ to the two copies of the derivative $\mu$. As 
$g_2$ has no rules, we get $d(f(r),f(s))=0$. Similarly, the rule defining $f_2$ applies operator $g_3$ in the target. The operator $g_3$ allows one to derive an unconditional move to the idle process $0$.
Hence, $d(f_2(r),f_2(s))=0$.

However, if the reactive behavior of the process associated to a source variable is tested by some premise, then the operator defined by this rule may discriminate states with different reactive behavior.
Consider the rules:
\begin{gather*}
	\SOSrule{x \trans[a] \mu}{f(x) \trans[a] g(\mu,\mu)}
\qquad\qquad
	\SOSrule{x_1 \trans[a] \mu_1 \quad x_2 \trans[a] \mu_2}{g(x_1,x_2) \trans[a] \delta_0}
\end{gather*}
These rules together with $R$ define $R_6$. 
We get $d(f(r),f(s))=1-(1-\epsilon)^2$ because the $a$-transition of term $r$ leads to a distribution where all states can perform the action $a$, but the $a$-transition of term $s$ leads to a distribution where only states with a total probability mass of $1-\epsilon$ can perform the action $a$.
\end{example}

We denote by $R_f$
those rules of $R$ that define the operator $f$.
We define by $\reactdist{f}{i} \in \{0,1\}$ the (reactive behavior) discriminating power of argument $i$ of
$f$. Formally, $\reactdist{f}{i} = 1$ if the source variable $x_i$ appears in a premise of some $\rho \in R_f$, i.e., if for some $\rho \in R_f$ there is a $t_k \trans[a_k] \mu_k \in \pprem{\rho}$ with $x_i \in \Var(t_k)$ or 
a ${t_l \ntrans[b_l]} \in {\nprem{\rho}}$ with $x_i \in \Var(t_l)$. Otherwise, $\reactdist{f}{i} = 0$. 
%
With $\Ninfty$ we denote 
$\N \cup \{\infty\}$, with the natural ordering extended by $n < \infty$ for each $n \in \N$, and the usual arithmetic extended for summation by $\infty + n = n + \infty = \infty + \infty = \infty$ for $n\ge 0$ and multiplication by $0 \cdot \infty = \infty \cdot 0 = 0$ and $n \cdot \infty = \infty \cdot n = \infty \cdot \infty = \infty$ for $n \ge 1$.

We quantify the 
expansivity power of operators $f \in \Sigma$ as least fixed point of a monotone function. Let $(\Sigma, \Act, R)$ with $\Sigma = (F, \rank)$ be a PTSS. We define a poset 
$\mathcal{S} = (S,\sqsubseteq)$ with $S = S_{\FF} \times S_{\VV}$, $S_{\FF} = F \times \N \to \Ninfty$, $S_{\VV}=(\openTerms\,\cup\,\openDTerms) \to ((\TVar\,\cup\,\PVar) \to \Ninfty)$, equipped with the point-wise partial order $(\GF, \GV) \sqsubseteq (\GF',\GV')$ iff $\GF(f,i) \le \GF'(f,i)$, for all $f \in F, i \in \N$, and $\GV(t)(\zeta) \le \GV'(t)(\zeta)$ for all $t \in \openTerms \cup \openDTerms, \zeta \in \TVar \cup \PVar$. Elements of $S$ are pairs of maps $(\GF,\GV)$. 
$\GF(f,i)$ denotes the 
expansivity power of argument $i$ in operator $f$, i.e., how much the operator $f$ multiplies the approximate bisimulation distance of argument $i$.
$\GV(t)(\zeta)$ defines the frequency of variable $\zeta \in \TVar \cup \PVar$ in the state or distribution term $t \in \openTerms \cup \openDTerms$ weighted by the 
expansivity power of the operators applied on top of $\zeta$.
$\mathcal{S}$ forms a complete lattice with bottom element $\bot$ and top element $\top$, defined by constant maps $\bot((f,i),(t,\zeta)) = (0,0)$ and  $\top((f,i),(t,\zeta))=(\infty,\infty)$ for each $f \in F$, $i \in \N$, $t \in \openTerms \cup \openDTerms$, $\zeta \in \TVar \cup \PVar$. 
\begin{proposition}\label{prop:s_complete_lattice}
	$\mathcal{S}$ is a complete lattice.	
\end{proposition}


The function $\functor: S \to S$ defined in Fig.~\ref{fig:functor_multiplicity_approx_bisim} computes in parallel the expansivity power of arguments of operators, and the multiplicities of variables in terms weighted by the expansivity power of the operators applied on top of them. The expansivity power $\GF'(f,i)$ of argument $i$ of operator $f$ is defined as the maximum expansivity power over each $f$-defining rule $\rho \in R_f$. For $\rho \in R_f$ the expansivity power is defined as the sum of the multiplicity of $x_i$ in the rule target $\target(\rho)$ and of the multiplicity of $x_i$ in some premise $t_k \trans[a] \mu_k \in \pprem{\rho}$ weighted by the multiplicity of the derivative $\mu_k$ in the rule target $\target(\rho)$. Note that source variables and derivatives in the rule target contribute equally to the expansivity power of an argument. The multiplicity $\GV'(t)(\zeta)$ of $\zeta$ in a state term $t$ counts the occurrences of variable $\zeta$ in $t$ and weights them by the expansivity power of the operators applied on top of $\zeta$. The multiplicity $\GV'(\theta)(\zeta)$ of $\zeta$ in a distribution term $\theta$ counts the occurrences of variable $\zeta$ in $\theta$ and weights them by the expansivity power of the operators applied on top of $\zeta$, but at least by the discriminating power of those operators. Note that the discriminating power of operators is considered only for distribution terms. To understand this, consider the reactive behavior of $\epsilon$-bisimilar state and distribution terms. 
For a state term $f(t_1,\ldots,t_{\rank(f)})$ we have that $\sigma(t_i) \bisim_{\epsilon_i} \sigma'(t_i)$ implies $\sigma(t_i) \trans[a]$ iff $\sigma(t_i') \trans[a]$ for each $a\in\Act$, i.e., $\sigma(t_i)$ and $\sigma(t'_i)$  agree on their immediate reactive behavior. However, for a distribution term $f(\theta_1,\ldots,\theta_{\rank(f)})$ we have that if $\sigma(\theta_i) \sim_{\epsilon_i} \sigma'(\theta_i)$ then $\sigma(\theta_i)$ and $\sigma'(\theta_i)$ may have states with different reactive behavior (cf. $R_6$ in Example~1). 

$\functor$ is order-preserving.
\remarkDG{Reviewer: Monotonicity is enough (continuity not required) since $S$ is a complete lattice.}
This ensures the existence and uniqueness of the least fixed point of $\functor$ by the Knaster-Tarski fixed point theorem.

\begin{proposition}\label{prop:fb_order_preserving}
	$\functor$ is order-preserving.
\end{proposition}

\begin{figure}[!t]
\begin{empheq}[box=\widefbox]{align*}
\\[-0.1cm]
&\!\!\!\!\!\!\!\!\!\!\!\!\!\!\!\!\!\!\!\!\!\!\!\!\!\!\!\!\!\!\!\text{Function }\functor: S \to S \text{ is defined by }\functor(\GF,\GV) = (\GF',\GV')\text{ with}\\[0.1cm]
\GF'(f,i) &= \sup_{\rho \in R_f} \biggl(
	\GV(\target(\rho))(x_i) + \displaystyle \!\!\!\!\!\!\sum_{t_k \trans[a_k] \mu_k \in \atop \pprem{\rho}} \!\!\!\!\!\GV(t_k)(x_i) \cdot \GV(\target(\rho))(\mu_k) 
	\biggr)\\[-0.1cm]
	\GV'(t)(\zeta) &= 
	\begin{cases}
		1 & \text{if } \zeta \in \TVar \text{ and }t=\zeta\\
		\displaystyle \sum_{i=1}^{\rank(f)} \left( \GF(f,i) \,\cdot\, \GV(t_i)(\zeta)\right)  \qquad\qquad\qquad & \text{if }t=f(t_1,\ldots,t_{\rank(f)})\\
		0 & \text{otherwise}
	\end{cases}\\[0.3cm]
  \GV'(\theta)(\zeta) &= 
	\begin{cases}
		1 				& \text{if } \theta = \zeta \\[0.2cm]
		\GV(t)(\zeta) 	& \text{if } \theta = \delta_t \\[0.2cm]
		\displaystyle\max_{i\in I} \Bigl(\GV(\theta_i)(\zeta)\Bigr) & \text{if } \theta=\sum_{i\in I} p_i \theta_i \\[0.2cm]
		\displaystyle \sum_{i=1}^{\rank(f)} \left( \max (\GF(f,i),\reactdist{f}{i} ) \,\cdot\, \GV(\theta_i)(\zeta)\right) & \text{if }\theta=f(\theta_1,\ldots,\theta_{\rank(f)}) \\[0.2cm]
		0 & \text{otherwise}
	\end{cases}\\[-0.2cm]
\end{empheq}
\caption{Function to quantify the approximate bisimulation multiplicity}
\label{fig:functor_multiplicity_approx_bisim}
\end{figure}

We denote the least fixed point of $\functor$ by $(\lfpF,\lfpT)$. We call $\lfpF(f,i)$ the expansivity power of argument $i$ of operator $f$, and $\lfpT(t)(\zeta)$ the weighted multiplicity of variable $\zeta$ in term $t$.
The expansivity power of $f$ allows us to derive an upper bound on the approximate bisimulation distance between terms $f(t_1,\ldots,t_{\rank(f)})$ and $f(t_1',\ldots,t_{\rank(f)}')$ 
expressed in relation to the approximate bisimulation distances $\epsilon_i$ between the arguments $t_i$ and $t_i'$.
\begin{definition}{\ \!\!\!\emph{(Expansivity bound)}\bf{.}\,} \label{def:upper_bound_expansion_approx_bisim}
The \emph{expansivity bound} $\expbound^f$ of operator $f \in \Sigma$ wrt.\ the approximate bisimulation distances $\epsilon_i$ of its arguments $i=1,\ldots,\rank(f)$ is defined by
\begin{equation*}
	\expbound^f(\epsilon_1,\ldots,\epsilon_{\rank(f)}) = 1 -\prod_{i=1}^{\rank(f)} (1-\epsilon_i)^{\lfpF(f,i)}
\end{equation*}
\end{definition}
%
Notice that $\expbound^f(\epsilon_1,\ldots,\epsilon_{\rank(f)}) = 0$ if $\epsilon_i=0$ for all arguments $i$ with $\lfpF(f,i) > 0$. In particular, we have $\expbound^f(\epsilon_1,\ldots,\epsilon_{\rank(f)}) = 0$ if all $\epsilon_i=0$. We call an argument $i$ of operator $f \in \Sigma$ (behavioral distance) \emph{absorbing} if $\lfpF(f,i) = 0$. 

We demonstrate first the application of the expansivity bound and prove later its correctness.

\begin{example}{\,(continued)\bf{.}\,} 
For the PTSS $(\Sigma,A,R_2)$ we have $\lfpF(f,1) = 2$ because $\lfpF(g,1)=\lfpF(g,2)=1$. Terms $r$ and $s$ with approximate bisimulation distance $d(r,s)=\epsilon$ agree by $1-\epsilon$ on their behavior. Thus, the pair of processes $(r,r)$ and $(s,s)$ agree by $(1-\epsilon)^2$ on their behavior. Hence, they disagree by $1-(1-\epsilon)^2$ on their behavior. This gives a behavioral distance of $d(f(r),f(s))=1-(1-\epsilon)^2$. 

We continue with PTSS $(\Sigma,A,R_3)$. For operator $h$ we have 
$\lfpF(h,1)=\lfpF(h,2)=\lfpF(h,3)=\lfpF(h,4)=1$,
 for 
$g$ we have $\lfpF(g,1)=\lfpF(g,2)=2$ and thus for 
$f$ we get $\lfpF(f,1)=4$. For PTSS $(\Sigma,A,R_4)$ the recursive definition of $f$ applied to the two occurrences of the derivative $\mu$ in the rule target gives $\lfpF(f,1)=\infty$. The (behavioral distance) absorbing effect of $f$ and $f_2$ in $(\Sigma,A,R_5)$ results in $\lfpF(f,1)=\lfpF(f_2,1)=\lfpF(g_2,1)=\lfpF(g_2,2)=\lfpF(g_3,1)=\lfpF(g_3,2)=0$. In $(\Sigma,\Act,R_6)$ the (reactive behavior) discriminating power $\reactdist{g}{1} = \reactdist{g}{2} = 1$ of operator $g$ leads to $\lfpF(f,1)=2$.
\end{example}

Now we can show that the approximate bisimulation distance between terms $f(t_1,\ldots,t_{\rank(f)})$ and $f(t_1',\ldots,t_{\rank(f)}')$ is bounded by the expansivity bound.

\begin{theorem}{\ \!\!\!\emph{(Expansivity bound of simple \ntmuft-PTSS)}\bf{.}\,} \label{thm:bounded_expansion_ntmuft_bisim}
Let $(\Sigma, \Act, R)$ be a stratifiable simple \ntmuft-PTSS. Then for each operator $f \in \Sigma$ we have
\[
	f(t_1,\dots,t_{\rank(f)}) \bisim_{\epsilon} f(t_1',\dots,t_{\rank(f)}') \quad\text{whenever}\quad t_i \bisim_{\epsilon_i} t_i' \ \text{ for } i=1,\ldots,\rank(f)
\]
with $\epsilon = \expbound^f(\epsilon_1,\ldots,\epsilon_{\rank(f)})$.
\end{theorem}
%
Theorem~\ref{thm:bounded_expansion_ntmuft_bisim} implies Theorem~\ref{thm:bisimilarity_as_congruence} by considering $\epsilon_i=0$ for all $i=1,\ldots,\rank(f)$ and exploiting that $\bisim_0$ is in fact the strict probabilistic bisimilarity.

Our target was to define the expansivity power $\lfpF(f)(i)$ of the argument $i$ of operator $f \in \Sigma$ in order to characterize the behavioral distance of terms with outermost function symbol $f$.
We conclude this section by outlining how the expansivity bound could be further refined. 
%
%
Sequential composition $\_\, ; \_$ is defined by the following rules \cite{DL12}
\begin{gather*}
	\SOSrule{x\trans[a]\mu}{{x; y}\trans[a] \mu; \delta_y}{\ a\neq\tick}
\qquad\qquad
	\SOSrule{x\trans[\tick]\mu \quad y\trans[a]\mu'}{{x; y}\trans[a]\mu'}
\end{gather*}
Action $\tick$ denotes successful termination. The expansivity power $\lfpF(;)(1) = \lfpF(;)(2) = 1$ gives an expansivity bound $\expbound^;(\epsilon_1,\epsilon_2) = 1-(1-\epsilon_1)(1-\epsilon_2)$. However, the sequential composition describes separate moves of either process $x$ or process $y$. Hence, the expansivity of $\_\, ; \_$ is actually bounded by $1-\min(1-\epsilon_1,1-\epsilon_2)$. In general, if multiple rules define an operator $f$, then the expansivity power and weighted multiplicity should be quantified per rule instead of per operator. In detail, the expansivity power $\lfpF(f)(i)$ should take a rule $\rho$ instead of $f$ as argument, and the weighted multiplicity $\lfpT(t,x)$ should take a tree of rules instead of term $t$ as argument. We leave this as future work.

\section{Specification of Non-expansive Process Combinators}\label{sec:non-exp}

Non-expansivity is the quantitative analogue of the congruence property of (strict) probabilistic bisimulation.
Intuitively, non-expansivity means that different processes are not more different 
when they are put in the same context. 

\begin{definition}{\ \!\!\!\emph{(Non-expansivity)}\bf{.}\,} \label{def:non_expansivity}
Let $(\closedTerms, \Act, {\trans}_{P})$ be the PTS induced by the PTSS $P=(\Sigma, \Act, R)$. An operator $f \in \Sigma$ is \emph{non-expansive} if 
\[
	f(t_1,\dots,t_{\rank(f)}) \bisim_{\epsilon} f(t_1',\dots,t_{\rank(f)}') \quad\text{whenever}\quad t_i \bisim_{\epsilon_i} t_i'\, \text{ for all }\, i=1,\ldots,\rank(f)
\]
with $\epsilon = \min(\sum_{i=1}^{\rank(f)} \epsilon_i,1)$. 
\end{definition}
We call $f$ \emph{expansive} if $f$ is not non-expansive. Argumentation for this linear upper bound and a discussion on alternative upper bounds like maximum norm or Euclidean norm can be found in \cite{Tin10}.

From the expansivity bound $\expbound^f$ (Definition~\ref{def:upper_bound_expansion_approx_bisim}) of operator $f \in \Sigma$ it follows that $f$ is non-expansive if 
$\lfpF(f,i) \le 1$ for all $i=1,\ldots,\rank(f)$. 
This yields the following rule format.

\begin{definition}{\ \!\!\!\emph{(\entmuft\ rule format)}\bf{.}\,} \label{def:entmuftRuleFormat}
A simple \ntmuft-rule $\rho$ is an \emph{\entmuft-rule} if for each $x_i \in \source(\rho)$ we have
\[
	\MVarMax(\target(\rho))(x_i) + \displaystyle \!\!\!\!\!\!\!\sum_{t_k \trans[a_k] \mu_k \in \atop \pprem{\rho}}\!\!\!\!\!\!\! \MVar(t_k)(x_i) \cdot \MVarMax(\target(\rho))(\mu_k) \le 1.
\]
A PTSS $P = (\Sigma,A,R)$ is in \entmuft\ format, \entmuft-PTSS for short, if all rules in $R$ are in the \entmuft\ rule format.
\end{definition}
%

\begin{theorem}{\ \!\!\!\emph{(Non-expansivity of \entmuft-PTSS)}\bf{.}\,} \label{thm:nonexp_entmuft_ptss}
Let $(\Sigma, \Act, R)$ be a stratifiable \entmuft-PTSS. Then all operators $f \in \Sigma$ are non-expansive.
\end{theorem}
The constraints of the \entmuft\ rule format are easy to verify. It suffices to count the occurrences of source variables and derivatives in the rule target. There is no need for recursive reasoning over other rules. We deliberately decided against the (slightly more general) rule format which could be given as simple \ntmuft-rules $\rho$ that define some operator $f\in\Sigma$ and for which the only requirement would be $\lfpF(f,i) \le 1$ for all $i=1,\ldots,\rank(f)$. We justify this by considering the extension of a PTSS $P=(\Sigma,\Act,R)$ to $P'=(\Sigma,\Act,R')$ with $R \subseteq R'$. If $P$ is in \entmuft\ format, then in order to decide if $P'$ is in \entmuft\ format only the rules in $R' \setminus R$ need to be verified wrt.\ the \entmuft\ format constraints. On the contrary, the generalized rule format would require that whenever a rule is added, all other rules are again validated with respect to the format constraints. For instance, consider the set of rules $R_6$ in Example~1. The rule defining operator $f$ alone would be non-expansive. However, by adding the rule defining operator $g$ (even though $g$ is non-expansive), operator $f$ becomes expansive. 

\section{Applications}\label{sec:applications}

The standard process combinators sequential composition, (probabilistic and non-probabilistic) choice, and (probabilistic and non-probabilistic) CCS and CSP like parallel composition \cite{Bar04,DL12} are all in the \entmuft-format. On the other hand, recursion and iteration operators may be expansive if they replicate (some of) their arguments. 
We consider the replication operator of $\pi$-calculus. The nondeterministic variant $!\_$ and the probabilistic variant $!^p\_$ with $p \in (0,1) \cap \Q$ are defined by the rules\cite{MS13}:
\begin{gather*}
	\SOSrule{x \trans[a] \mu}{!x \trans[a] \mu \parallel \delta_{!x}}
\qquad
	\SOSrule{x \trans[a] \mu}{!^px \trans[a] \mu \oplus_p (\mu \parallel \delta_{!^px})} 
\qquad
%
	\SOSrule{x_1 \trans[a] \mu_1 \quad x_2 \trans[a] \mu_2}{x_1 \parallel x_2 \trans[a] \mu_1 \parallel \mu_2}
\end{gather*}
The first two rules defining both variants of the replication operator are not in \entmuft-format. The expansivity power of both operators is unbounded with $\lfpF(!)(1)=\lfpF(!^p)(1)=\infty$. Hence, both operators are expansive. However, if the synchronous parallel composition defined in the third rule above is replaced by the non-communicating asynchronous parallel composition, then both variants of the replication operator would become non-expansive. 

\begin{table}[t!]
	\qquad\quad\qquad\ \ 
	\subfloat{\vtop{\vskip0pt\hbox{
	\begin{tikzpicture}
		\node (r) at (0,0) {$r$} ;
		\node (pi) at ($ (r) - (0,1) $) {$\pi_r$};

		\path[->] (r)  [bend right = 80,distance=1.25cm] edge node [left] {{\scriptsize $a$}} (pi);
		\path[->] (r)  [bend left = 80,distance=1.25cm] edge node [right] {{\scriptsize $b$}} (pi);
		\path[->] (pi) [dotted] edge node [right] {{\scriptsize $1.0$}} (r);
	\end{tikzpicture}
	}}}
	\quad\ \ 
	\subfloat{\vtop{\vskip0pt\hbox{
	\begin{tikzpicture}
		\node (s) at (0,0) {$s$} ;
		\node (pi) at ($ (s) - (0,1) $) {$\pi_s$};
		\node (s') at ($ (pi) - (0,1) $) {$s'$};

		\path[->] (s)  [bend right = 80,distance=1.25cm] edge node [left] {{\scriptsize $a$}} (pi);
		\path[->] (s)  [bend left = 80,distance=1.5cm] edge node [right] {{\scriptsize $b$}} (pi);
		\path[->] (pi) [dotted] edge node [right] {{\scriptsize $1.0-\epsilon$}} (s);		
		\path[->] (pi) [dotted] edge node [right] {{\scriptsize $\epsilon$}} (s');
		\path[->] (s')  [bend right = 80,distance=1.25cm] edge node [right] {{\scriptsize $c$}} (pi);
	\end{tikzpicture}
	}}}
	\quad\ \ 
	\subfloat{\vtop{\vskip0pt\hbox{
	\begin{tikzpicture}
		\node at (0,0) {Common rule};
		\node at (0,-1) {$\displaystyle{
			\SOSrule{x_1 \trans[a] \mu_1  \quad  x_2 \trans[a] \mu_2}
					{x_1 \parallel x_2 \trans[a]  \mu_1 \parallel \mu_2}
			}$};
	\end{tikzpicture}
	}}}
\vspace{0cm}\\
\begin{center}
\begin{tabular}{|@{\hskip 0.4cm}l@{\hskip 0.4cm}|@{\hskip 0.4cm}c@{\hskip 0.4cm}|}
\hline 
\hspace{0.5cm}Description & Rule \\ 
\hline
& \\&\vspace{-0.5cm}\\
$1\hspace{0.15cm}\begin{array}{l}
\text{Non-linearity of the rule target}\\
\text{wrt.\ a source variable}
\end{array}$ & $\displaystyle{
		\SOSrule{x \trans[a] \mu}
    			{f(x) \trans[a]  \delta_x \parallel \delta_x}
		}$ \\&\vspace{-0.2cm}\\
$2\hspace{0.15cm}\begin{array}{l}
\text{Non-linearity of the rule target}\\
\text{wrt.\ a derivative}
\end{array}$ & $\displaystyle{
		\SOSrule{x \trans[a] \mu}
	    		{f(x) \trans[a]  \mu \parallel \mu}
		}$ \\&\vspace{-0.2cm}\\
$3\hspace{0.15cm}\begin{array}{l}
\text{Non-linearity of a state term}\\
\text{in the rule target}
\end{array}$ & $\displaystyle{
		\SOSrule{x \trans[a] \mu}
	    		{f(x) \trans[a]  \delta_{x \parallel x}}
		}$ \\&\vspace{-0.2cm}\\
$4\hspace{0.15cm}\begin{array}{l}
\text{Non-linearity of a term in a}\\
\text{premise}
\end{array}$ & $\displaystyle{
		\SOSrule{x \parallel x \trans[a] \mu}
				{f(x) \trans[a]  \mu}
		}$ \\&\vspace{-0.2cm}\\
$5\hspace{0.15cm}\begin{array}{l}
\text{Multiple derivatives of a source}\\
\text{variable in the rule target}
\end{array}$ & $\displaystyle{
	\SOSrule{x \trans[a] \mu_1  \quad   x \trans[b] \mu_2}
		 	{f(x) \trans[a]  \mu_1 \parallel \mu_2}
	}$ \\&\vspace{-0.2cm}\\
$6\hspace{0.15cm}\begin{array}{l}
\text{Source and derivative in the}\\
\text{rule target}
\end{array}$ & $\displaystyle{
		\SOSrule{x \trans[a] \mu}
	    		{f(x) \trans[a]  \delta_x \parallel \mu}
		}$ \\&\vspace{-0.2cm}\\
$7\hspace{0.15cm}\begin{array}{l}
\text{Multiple derivatives weighted by}\\
\text{convex combination in rule target}
\end{array}$ & $\displaystyle{
		\SOSrule{x \trans[a] \mu}
	    		{f(x) \trans[a]  [0.5] \mu \parallel \mu \oplus [0.5]\delta_{s'}}
		}$ \\&\vspace{-0.2cm}\\
$8\hspace{0.15cm}\begin{array}{l}
\text{Lookahead by existential test}\\
\text{in quantitative premise}
\end{array}$ & $\displaystyle{
		\SOSrule{x \trans[a] \mu_1 \quad \mu_1(y)>0 \quad y \trans[c] \mu_2}
	    		{f(x) \trans[a]  \mu_2}
	}$ \\&\vspace{-0.2cm}\\
$9\hspace{0.15cm}\begin{array}{l}
\text{Lookahead by universal test}\\
\text{in quantitative premise}
\end{array}$ & $\displaystyle{
		\SOSrule{x \trans[a] \mu \quad \mu(Y) \ge 1 \quad \{ y \trans[a] \mu_y \mid y\in Y \}}
				{f(x) \trans[a]  \delta_r}
		}$ \vspace{-0.2cm}\\&\\
\hline
\end{tabular}
\end{center}
\vspace{0cm}
\caption{SOS rules that specify expansive operators}\label{tab:counterexample_rule_restrictions_entmufxt}
\end{table}

We summarize the structural patterns of rules that may lead to expansive behavior in Table~\ref{tab:counterexample_rule_restrictions_entmufxt}. None of these rules is in the \entmuft\ format. For cases 1 to 7 the expansivity power of $f$ is $\lfpF(f)(1) = 2$ and, therefore, the expansivity bound is $\expbound^f(\epsilon) = d(f(r),f(s)) = 1 - (1-\epsilon)^2$.
Cases 8 and 9 indicate that lookahead cannot be admitted and we need to employ simple \ntmuft-rules (Definition~\ref{def:simple_ntmuft}) instead of \ntmuft-rules \cite{LGD12}. The expressions $\mu_1(y)>0$ and $\mu_1(Y) \ge 1$ (with $y\in\TVar$, $Y \subseteq \TVar$) are quantitative premises as introduced by the \ntmufxt\ format \cite{DL12}.
As argued above, $\epsilon$-bisimilar instances may have states with different reactive behavior. 
For instance, in case 8, while distributions $\pi_r$ and $\pi_s$ are $\epsilon$-bisimilar, only $\pi_s$ has in its support a state that can perform a $c$-move. Similarly, in case 9, only for $\pi_r$ we have that all states in the support can perform an $a$-move.

We conjecture that a notion of \emph{non-expansivity up to $\epsilon$} for some $\epsilon \in [0,1]$ (bounded non-expansivity) would allow for limited lookahead. An operator is non-expansive up to $\epsilon$ if it is non-expansive whenever its arguments have an approximate bisimulation distance of at most $\epsilon$. In this case, quantitative premises $\mu(Y) \ge p$ that measure the probability of $Y$ and test against the boundary $p$ could be allowed, if $p$ is in the interval $p \in [\epsilon,1-\epsilon]$. On the other hand, quantitative premises with $p < \epsilon$ or $p > 1-\epsilon$ cannot be permitted because they allow for lookahead with respect to probabilistic choices that do not mimic each other's reactive behavior.


The expansivity bound (Definition~\ref{def:upper_bound_expansion_approx_bisim}) allows rule formats to be derived for alternative compositionality requirements. For instance, consider an $n$-ary process combinator $\otimes$ with the compositionality requirements that the approximate bisimulation distance of the combined processes should not depend on the approximate bisimulation distance of processes at some argument $i \in \{1,\ldots,n\}$. From the expansivity bound we derive that either argument $i$ of operator $\otimes$ is behavioral distance absorbing ($\lfpF(\otimes)(i) = 0$), or the application context guarantees that processes for argument $i$ are strictly bisimilar ($\epsilon_i = 0$). 

The non-expansivity requirement (Definition~\ref{def:non_expansivity}) is in fact the Manhattan norm, or more general the $p$-norm $(\sum_{i=1}^{\rank(\otimes)}\epsilon_i^p)^{1/p}$ with $p=1$. Consider the alternative compositionality requirement that the expansivity of a process combinator $\otimes$ should be bounded by the $p$-norm with $p>1$ (which includes the Euclidean norm by $p=2$ and the maximum norm by $p \to \infty$). From the expansivity bound we derive that $\lfpF(\otimes)(i) = 1$ for at most one argument $i$, and all other arguments $j \ne i$ are behavioral distance absorbing with $\lfpF(\otimes)(j) = 0$.

\section{Conclusion and Future Work}\label{sec:conclusion}

We studied structural specifications of probabilistic processes that are robust with respect to bounded implementation and measurement errors of probabilistic behavior. We provided for each process combinator an upper bound on the distance between the combined processes using the structural specification of the process combinator (Theorem~\ref{thm:bounded_expansion_ntmuft_bisim}). We derived an appropriate rule format that guarantees non-expansivity (standard compositionality requirement) of process combinators (Theorem~\ref{thm:nonexp_entmuft_ptss}). All standard process algebraic operators are compatible for approximate reasoning and satisfy the rule format, except operators which replicate processes and combine them by synchronous parallel composition. We exemplified how rule formats for non-standard compositionality requirements can be derived. 

Our work can be extended in several directions. In Section~\ref{sec:aptss} and Section~\ref{sec:applications} we sketched already how the expansivity bound can be further refined and how a restricted form of lookahead in the rules specifying the process combinators could be admitted. The techniques and results developed in this paper for approximate bisimulation can be carried over to bisimulation metrics. 
Initial work in this direction suggests 
that the \entmuft\ format presented in this paper ensures also non-expansivity for the bisimulation metric based on the Kantorovich and Hausdorff metric. Moreover, for the bisimulation metric the rule format can be further generalized because in this case convex combinations weigh the distance and multiplicity of processes (unlike approximate bisimilarity, see case~7 of Table~\ref{tab:counterexample_rule_restrictions_entmufxt}). Furthermore, we will investigate the expansivity of process combinators and rule formats for variants of bisimulation metrics and $\epsilon$-bisimulation that discount the influence of future transitions \cite{DGJP04,TDZ11}.

\paragraph{Acknowledgements} We are grateful to Jos\'ee Desharnais for discussions on $\epsilon$-bisimulation, Matteo Mio for discussions on approximate semantics of structurally defined probabilistic systems, and Wan Fokkink and David Williams for feedback on earlier versions of this paper. Furthermore, we thank the anonymous referees for thorough reviews and very helpful comments.

\bibliographystyle{eptcs}
\bibliography{concur}

\begin{thebibliography}{10}
\providecommand{\bibitemdeclare}[2]{}
\providecommand{\surnamestart}{}
\providecommand{\surnameend}{}
\providecommand{\urlprefix}{Available at }
\providecommand{\url}[1]{\texttt{#1}}
\providecommand{\href}[2]{\texttt{#2}}
\providecommand{\urlalt}[2]{\href{#1}{#2}}
\providecommand{\doi}[1]{doi:\urlalt{http://dx.doi.org/#1}{#1}}
\providecommand{\bibinfo}[2]{#2}

\bibitemdeclare{inproceedings}{BM12}
\bibitem{BM12}
\bibinfo{author}{Giorgio \surnamestart Bacci\surnameend} \&
  \bibinfo{author}{Marino \surnamestart Miculan\surnameend}
  (\bibinfo{year}{2012}): \emph{\bibinfo{title}{Structural Operational
  Semantics for Continuous State Probabilistic Processes}}.
\newblock In: {\sl \bibinfo{booktitle}{Proc.~CMCS'12}}, {\sl
  \bibinfo{series}{LNCS}} \bibinfo{volume}{7399},
  \bibinfo{publisher}{Springer}, pp. \bibinfo{pages}{71--89},
  \doi{10.1007/978-3-642-32784-1\_5}.

\bibitemdeclare{phdthesis}{Bar04}
\bibitem{Bar04}
\bibinfo{author}{Falk \surnamestart Bartels\surnameend} (\bibinfo{year}{2004}):
  \emph{\bibinfo{title}{On Generalised Coinduction and Probabilistic
  Specification Formats}}.
\newblock Ph.D. thesis, \bibinfo{school}{VU University Amsterdam}.

\bibitemdeclare{article}{BW05}
\bibitem{BW05}
\bibinfo{author}{Franck \surnamestart van Breugel\surnameend} \&
  \bibinfo{author}{James \surnamestart Worrell\surnameend}
  (\bibinfo{year}{2005}): \emph{\bibinfo{title}{A Behavioural Pseudometric for
  Probabilistic Transition Systems}}.
\newblock {\sl \bibinfo{journal}{Theor. Comput. Sci.}}
  \bibinfo{volume}{331}(\bibinfo{number}{1}), pp. \bibinfo{pages}{115--142},
  \doi{10.1016/j.tcs.2004.09.035}.

\bibitemdeclare{inproceedings}{DL12}
\bibitem{DL12}
\bibinfo{author}{Pedro~R. \surnamestart D'Argenio\surnameend} \&
  \bibinfo{author}{Matias~David \surnamestart Lee\surnameend}
  (\bibinfo{year}{2012}): \emph{\bibinfo{title}{Probabilistic Transition System
  Specification: Congruence and Full Abstraction of Bisimulation}}.
\newblock In: {\sl \bibinfo{booktitle}{Proc.~FoSSaCS'12}}, {\sl
  \bibinfo{series}{LNCS}} \bibinfo{volume}{7213},
  \bibinfo{publisher}{Springer}, pp. \bibinfo{pages}{452--466},
  \doi{10.1007/978-3-642-28729-9\_30}.

\bibitemdeclare{article}{DGJP03}
\bibitem{DGJP03}
\bibinfo{author}{Jos\'ee \surnamestart Desharnais\surnameend},
  \bibinfo{author}{Vineet \surnamestart Gupta\surnameend},
  \bibinfo{author}{Radha \surnamestart Jagadeesan\surnameend} \&
  \bibinfo{author}{Prakash \surnamestart Panangaden\surnameend}
  (\bibinfo{year}{2003}): \emph{\bibinfo{title}{Approximating Labelled Markov
  Processes}}.
\newblock {\sl \bibinfo{journal}{Information and Computation}}
  \bibinfo{volume}{184}(\bibinfo{number}{1}), pp. \bibinfo{pages}{160 -- 200},
  \doi{10.1016/S0890-5401(03)00051-8}.

\bibitemdeclare{article}{DGJP04}
\bibitem{DGJP04}
\bibinfo{author}{Jos\'ee \surnamestart Desharnais\surnameend},
  \bibinfo{author}{Vineet \surnamestart Gupta\surnameend},
  \bibinfo{author}{Radha \surnamestart Jagadeesan\surnameend} \&
  \bibinfo{author}{Prakash \surnamestart Panangaden\surnameend}
  (\bibinfo{year}{2004}): \emph{\bibinfo{title}{Metrics for Labelled Markov
  Processes}}.
\newblock {\sl \bibinfo{journal}{Theor. Comput. Sci.}}
  \bibinfo{volume}{318}(\bibinfo{number}{3}), pp. \bibinfo{pages}{323--354},
  \doi{10.1016/j.tcs.2003.09.013}.

\bibitemdeclare{inproceedings}{DLT08}
\bibitem{DLT08}
\bibinfo{author}{Jos\'ee \surnamestart Desharnais\surnameend},
  \bibinfo{author}{Francois \surnamestart Laviolette\surnameend} \&
  \bibinfo{author}{Mathieu \surnamestart Tracol\surnameend}
  (\bibinfo{year}{2008}): \emph{\bibinfo{title}{Approximate Analysis of
  Probabilistic Processes: Logic, Simulation and Games}}.
\newblock In: {\sl \bibinfo{booktitle}{Proc.~QEST'08}},
  \bibinfo{organization}{IEEE}, pp. \bibinfo{pages}{264--273},
  \doi{10.1109/QEST.2008.42}.

\bibitemdeclare{inproceedings}{GJS90}
\bibitem{GJS90}
\bibinfo{author}{Alessandro \surnamestart Giacalone\surnameend},
  \bibinfo{author}{Chi-Chang \surnamestart Jou\surnameend} \&
  \bibinfo{author}{Scott~A. \surnamestart Smolka\surnameend}
  (\bibinfo{year}{1990}): \emph{\bibinfo{title}{Algebraic Reasoning for
  Probabilistic Concurrent Systems}}.
\newblock In: {\sl \bibinfo{booktitle}{Proc.~IFIP TC2 Working Conf. on Prog.
  Concepts and Methods}}, pp. \bibinfo{pages}{443--458}.

\bibitemdeclare{article}{vG04}
\bibitem{vG04}
\bibinfo{author}{Rob~J. \surnamestart van Glabbeek\surnameend}
  (\bibinfo{year}{2004}): \emph{\bibinfo{title}{The Meaning of Negative
  Premises in Transition System Specifications {II}}}.
\newblock {\sl \bibinfo{journal}{J. Log. Algebr. Program.}}
  \bibinfo{volume}{60-61}, pp. \bibinfo{pages}{229--258},
  \doi{10.1016/j.jlap.2004.03.007}.

\bibitemdeclare{article}{Gro93}
\bibitem{Gro93}
\bibinfo{author}{Jan~Friso \surnamestart Groote\surnameend}
  (\bibinfo{year}{1993}): \emph{\bibinfo{title}{Transition System
  Specifications with Negative Premises}}.
\newblock {\sl \bibinfo{journal}{Theor. Comput. Sci.}}
  \bibinfo{volume}{118}(\bibinfo{number}{2}), pp. \bibinfo{pages}{263--299},
  \doi{10.1016/0304-3975(93)90111-6}.

\bibitemdeclare{article}{GV92}
\bibitem{GV92}
\bibinfo{author}{Jan~Friso \surnamestart Groote\surnameend} \&
  \bibinfo{author}{Frits \surnamestart Vaandrager\surnameend}
  (\bibinfo{year}{1992}): \emph{\bibinfo{title}{Structured Operational
  Semantics and Bisimulation as a Congruence}}.
\newblock {\sl \bibinfo{journal}{Inf. Comput.}} \bibinfo{volume}{100}, pp.
  \bibinfo{pages}{202--260}, \doi{10.1016/0890-5401(92)90013-6}.

\bibitemdeclare{incollection}{LT05}
\bibitem{LT05}
\bibinfo{author}{Ruggero \surnamestart Lanotte\surnameend} \&
  \bibinfo{author}{Simone \surnamestart Tini\surnameend}
  (\bibinfo{year}{2005}): \emph{\bibinfo{title}{Probabilistic Congruence for
  Semistochastic Generative Processes}}.
\newblock In: {\sl \bibinfo{booktitle}{Proc.~FoSSaCS'05}}, {\sl
  \bibinfo{series}{LNCS}} \bibinfo{volume}{3441},
  \bibinfo{publisher}{Springer}, pp. \bibinfo{pages}{63--78},
  \doi{10.1007/978-3-540-31982-5\_4}.

\bibitemdeclare{article}{LT09}
\bibitem{LT09}
\bibinfo{author}{Ruggero \surnamestart Lanotte\surnameend} \&
  \bibinfo{author}{Simone \surnamestart Tini\surnameend}
  (\bibinfo{year}{2009}): \emph{\bibinfo{title}{Probabilistic Bisimulation as a
  Congruence}}.
\newblock {\sl \bibinfo{journal}{ACM TOCL}} \bibinfo{volume}{10}, pp.
  \bibinfo{pages}{1--48}, \doi{10.1145/1462179.1462181}.

\bibitemdeclare{article}{LS91}
\bibitem{LS91}
\bibinfo{author}{Kim~G. \surnamestart Larsen\surnameend} \&
  \bibinfo{author}{Arne \surnamestart Skou\surnameend} (\bibinfo{year}{1991}):
  \emph{\bibinfo{title}{Bisimulation Through Probabilistic Testing}}.
\newblock {\sl \bibinfo{journal}{Inf. Comput.}} \bibinfo{volume}{94}, pp.
  \bibinfo{pages}{1--28}, \doi{10.1016/0890-5401(91)90030-6}.

\bibitemdeclare{inproceedings}{LGD12}
\bibitem{LGD12}
\bibinfo{author}{Matias~David \surnamestart Lee\surnameend},
  \bibinfo{author}{Daniel \surnamestart Gebler\surnameend} \&
  \bibinfo{author}{Pedro~R. \surnamestart D'Argenio\surnameend}
  (\bibinfo{year}{2012}): \emph{\bibinfo{title}{Tree Rules in Probabilistic
  Transition System Specifications with Negative and Quantitative Premises}}.
\newblock In: {\sl \bibinfo{booktitle}{Proc.~EXPRESS/SOS'12}}, {\sl
  \bibinfo{series}{EPTCS}}~\bibinfo{volume}{89}, pp. \bibinfo{pages}{115--130},
  \doi{10.4204/EPTCS.89.9}.

\bibitemdeclare{inproceedings}{MS13}
\bibitem{MS13}
\bibinfo{author}{Matteo \surnamestart Mio\surnameend} \& \bibinfo{author}{Alex
  \surnamestart Simpson\surnameend} (\bibinfo{year}{2013}):
  \emph{\bibinfo{title}{A {P}roof {S}ystem for {C}ompositional {V}erification
  of {P}robabilistic {C}oncurrent {P}rocesses}}.
\newblock In: {\sl \bibinfo{booktitle}{Proc.~FoSSaCS'13}}, {\sl
  \bibinfo{series}{LNCS}} \bibinfo{volume}{7794},
  \bibinfo{publisher}{Springer}, pp. \bibinfo{pages}{161--176},
  \doi{10.1007/978-3-642-37075-5\_11}.

\bibitemdeclare{techreport}{Plo04}
\bibitem{Plo04}
\bibinfo{author}{Gordon \surnamestart Plotkin\surnameend}
  (\bibinfo{year}{1981}): \emph{\bibinfo{title}{A Structural Approach to
  Operational Semantics}}.
\newblock \bibinfo{type}{Report} \bibinfo{number}{DAIMI FN-19},
  \bibinfo{institution}{Aarhus University}.

\bibitemdeclare{phdthesis}{Seg95a}
\bibitem{Seg95a}
\bibinfo{author}{Roberto \surnamestart Segala\surnameend}
  (\bibinfo{year}{1995}): \emph{\bibinfo{title}{Modeling and Verification of
  Randomized Distributed Real-Time Systems}}.
\newblock Ph.D. thesis, \bibinfo{school}{MIT}.

\bibitemdeclare{inproceedings}{Tin08}
\bibitem{Tin08}
\bibinfo{author}{Simone \surnamestart Tini\surnameend} (\bibinfo{year}{2008}):
  \emph{\bibinfo{title}{Non Expansive $\epsilon$-bisimulations}}.
\newblock In: {\sl \bibinfo{booktitle}{Proc.~AMAST'08}}, {\sl
  \bibinfo{series}{LNCS}} \bibinfo{volume}{5140},
  \bibinfo{publisher}{Springer}, pp. \bibinfo{pages}{362--376},
  \doi{10.1007/978-3-540-79980-1\_27}.

\bibitemdeclare{article}{Tin10}
\bibitem{Tin10}
\bibinfo{author}{Simone \surnamestart Tini\surnameend} (\bibinfo{year}{2010}):
  \emph{\bibinfo{title}{Non-expansive $\epsilon$-bisimulations for
  Probabilistic Processes}}.
\newblock {\sl \bibinfo{journal}{Theoret. Comput. Sci.}} \bibinfo{volume}{411},
  pp. \bibinfo{pages}{2202--2222}, \doi{10.1016/j.tcs.2010.01.027}.

\bibitemdeclare{phdthesis}{Trac10}
\bibitem{Trac10}
\bibinfo{author}{Mathieu \surnamestart Tracol\surnameend}
  (\bibinfo{year}{2010}): \emph{\bibinfo{title}{Approximate Verification of
  Probabilistic Systems}}.
\newblock \bibinfo{type}{Dissertation}, \bibinfo{school}{{LRI, Universit\'e
  Paris-Sud}}.

\bibitemdeclare{inproceedings}{TDZ11}
\bibitem{TDZ11}
\bibinfo{author}{Mathieu \surnamestart Tracol\surnameend},
  \bibinfo{author}{Jos\'ee \surnamestart Desharnais\surnameend} \&
  \bibinfo{author}{Abir \surnamestart Zhioua\surnameend}
  (\bibinfo{year}{2011}): \emph{\bibinfo{title}{Computing Distances between
  Probabilistic Automata}}.
\newblock In: {\sl \bibinfo{booktitle}{Proc.~QAPL'11}}, {\sl
  \bibinfo{series}{EPTCS}}~\bibinfo{volume}{57}, pp. \bibinfo{pages}{148--162},
  \doi{10.4204/EPTCS.57.11}.

\end{thebibliography}

\end{document}